\documentclass[useAMS,usenatbib]{mn2e} 
\usepackage[english]{babel}
\usepackage{journals}
\usepackage[pdftex]{graphicx,color} 
\usepackage[latin1]{inputenc}
\usepackage{graphics} 
\usepackage{amsfonts} 
\usepackage{amsmath}
\usepackage{multicol} 
\usepackage{layout} 
\usepackage{amssymb}
\usepackage[colorlinks=true,pdfstartview=FitV,linkcolor=red,citecolor=blue,urlcolor=magenta]{hyperref}

\title[Universality of dark matter haloes shape]{Universality of dark matter haloes shape over six decades in mass: Insights from the Millennium XXL and SBARBINE simulations}
\author[Bonamigo et al. 2014]
{\parbox{\textwidth}{
	Mario Bonamigo$^{1}$\thanks{E-mail: \href{mailto:mario.bonamigo@lam.fr} {mario.bonamigo@lam.fr}},
	Giulia Despali$^{2}$, Marceau Limousin$^{1}$, Raul Angulo$^{3}$, Carlo Giocoli$^{4,5,6}$, 
        Genevi\`eve Soucail$^{7,8}$} \\ \\ 
$^{1}$ Aix Marseille Universit\'e, CNRS, LAM (Laboratoire
d'Astrophysique de Marseille) UMR 7326, 13388, Marseille, France \\
$^{2}$ Dipartimento di Fisica e Astronomia, Universit\`a degli Studi di Padova,
vicolo dell'Osservatorio 3, 35122, Padova, Italy \\
$^{3}$ Centro de Estudios de F\'isica del Cosmos de Arag\'on (CEFCA),
Plaza San Juan 1, Planta-2, 44001, Teruel, Spain \\
 $^{4}$  Dipartimento  di Fisica  e  Astronomia,  Alma Mater  Studiorum
  Universit\`{a} di Bologna, viale Berti Pichat, 6/2, 40127
  Bologna, Italy\\
  $^{5}$ INAF - Osservatorio Astronomico di
  Bologna, via Ranzani 1, 40127, Bologna, Italy \\
  $^{6}$  INFN -  Sezione di  Bologna,  viale Berti  Pichat 6/2,  40127,
  Bologna, Italy \\
$^{7}$ Universit\'e de Toulouse, UPS-OMP,  Institut de Recherche en Astrophysique et Plan\'etologie (IRAP), Toulouse, France \\
$^{8}$ CNRS, IRAP, 14 Avenue Edouard Belin, F-31400 Toulouse, France
}

\defcitealias{Planckxvi}{Planck Collaboration XIV} 	  	 

\begin{document}
\date{}
\maketitle
\label{firstpage}
\pagerange{\pageref{firstpage}--\pageref{lastpage}} \pubyear{2014}

\begin{abstract}
  For the last 30 years many observational and theoretical
  evidences  have shown  that  galaxy  clusters are  not  spherical
  objects,  and that  their  shape  is much  better  described by  a
  triaxial  geometry.  With  the  advent of  multi-wavelength data  of
  increasing  quality, triaxial investigations  of galaxy  clusters is
  gathering a growing interest from the community, especially in the time of ``precision cosmology''.

  In this work,  we aim  to
  provide  the  first  statistically  significant predictions  in  the
  unexplored mass  range above $3\times10^{14}~$M$_\odot h^{-1}$,  using haloes from
  two redshifts ($z=0$ and $z=1$) of the  Millennium XXL simulation.
  The size of this cosmological  dark  matter   only  simulation ($4.1~$Gpc) allows the formation of a statistically  significant
  number of massive cluster scale haloes ($\approx 500$  with  M$>$
  2$\times$ 10$^{15}~$M$_\odot h^{-1}$, and 780\,000 with M$>$ 10$^{14}~$M$_\odot h^{-1}$).
  Besides, we aim to extend  this investigation to
  lower masses in order to look for universal predictions across
  nearly six orders  of magnitude in mass, from  $10^{10}$ to almost $10^{16}~$M$_\odot h^{-1}$.   For  this purpose  we  use   the  SBARBINE
  simulations, allowing to model  haloes of masses starting from $\approx
  10^{10}~$M$_\odot h^{-1}$.
  We use an elliptical overdensity method to select haloes and compute
  the  shapes of the unimodal ones (approximately $50$ per cent), while we discard the
  unrelaxed.

  The minor to major and intermediate to major axis ratio are found to
  be  well described  by  simple functional  forms.   For a  given  mass we can  fully
  characterize the shape of a halo and give predictions about the
  distribution of axis ratios for a given cosmology and redshift.
  Moreover, these results are in some disagreement with the findings of \citet{Jing2002} which are widely used in the community even though they have to be extrapolated far beyond their original mass range.
   This ``recipe'' is made available to
  the community in this paper and in a dedicated web page.

\end{abstract}


\section{Introduction}
Spectroscopic galaxy redshift surveys and numerical N-body simulations
have revealed a large scale distribution of matter in the Universe featuring
a complex network of interconnected filamentary galaxy associations.
Vertices, i.e. intersections among the filaments, correspond to the very
dense compact nodes within this
\emph{cosmic web} where one can find massive galaxy clusters.

These objects have been first assigned a spherical geometry, being the 
easiest way to characterize an shape in three dimensions; at the time this fitted the available data well enough.
Nowadays, with the advent of multi-wavelength data of increasing quality, 
there is a growing interest from the community to go beyond the 
spherical assumption, which is inaccurate and misleading.

There is much observational evidence for clusters not being spherical
objects, from the non-circular projection of various probes:
optical, density maps of cluster galaxies \citep{carter80,bingelli82};
X-ray, surface brightness maps \citep{fabricant84,buote92,buote96,kawahara,lau2012};
Sunyaev Zel'dovich pressure maps \citep{sayers2011a};
strong gravitational lensing \citep{soucail87}, and
weak gravitational lensing \citep{evans09,oguri10,oguri11}.

Recently, the azimuthal variation of galaxy kinematics has been detected for the first time
in a stacked sample
of 1\,743 galaxy clusters from the SDSS \citep{skielboe}. They find that the line of sight
velocity dispersion of galaxies lying along the major axis of the central galaxy is larger
than those that lie along the minor axis, a detection providing further evidence
for the asphericity of galaxy clusters.

On the numerical side, haloes forming in cosmological simulations have been found
to be triaxial in shape, with a preference for prolateness over oblateness \citep{frenk88,dubinski91,warren92,cole96,Jing2002,hopkins05,bailin05,kasunevrard05,paz06,allgood,bett07,munozcuartas,phoenix,Schneider2012,Despali2013}.
These simulations also predict an evolution of the shape with mass and redshift:
low mass haloes appear more spherical than high mass haloes,
essentially because high mass haloes have formed
later on \citep{Despali2014}.

Finally, it can be shown \citep{doroshkevich} that triaxial
collapse is a straightforward prediction of structure growth
driven by self-gravity of Gaussian density fluctuations.

Therefore, the triaxial framework, though still being an
approximation, encapsulates the shapes much more accurately
than the spherical counterpart.

Besides, it has been shown that the cluster properties
(mass, concentration parameter, slope of the inner
dark matter density profile, strong lensing cross section) can differ significantly
depending on the shape assumed in the analysis
\citep[see, e.g. the discussion in][regarding 
Abell~1689]{Limousin2013}\citep[see also][]{Giocoli2012a,Giocoli2012b}.
Even the galaxy correlation function can be affected by wrong
assumptions on the triaxiality of haloes \citep{Daalen2012}.

Since these properties constitute the key ingredients of
some cosmological tests, this suggests that in the 
road map of ``precision cosmology'' with galaxy clusters, triaxial modelling is the next milestone.

In this paper, we aim to characterize the shape of numerically simulated clusters,
described within a triaxial framework.
Apart from the three Euler angles, a triaxial geometry is characterized by three axes ($a<b<c$), hence two axial ratios: minor to major ($s=a/c$ in the following) and intermediate to major ($q=b/c$).

Shape  of triaxial haloes  have been  investigated theoretically  in a
number of works which aim  to characterise the dependence of shapes on
mass, redshift, radius and so on.  Most of the works agree on the fact
that massive haloes are on average more elongated than low mass haloes
\citep{Jing2002,Allgood2006,munozcuartas,Despali2013,Despali2014},
since they form  at later times and thus still  retain memory of their
original shape which is influenced by the direction of the surrounding
filaments or of the last major merger; moreover, shapes depend also on
redshift with  haloes of  all masses having  on average  smaller axial
ratios in  the past  even if the  rank in  mass is maintained  at all
times    \citep{munozcuartas,Despali2014}.     Other    works    have
investigated halo shapes as a  function of radius, measuring the axial
ratios  of shells  at  different  distances from  the  centre and  the
alignment               between               the               shells
\citep{warren92,Jing2002,bailin05,Allgood2006,Schneider2012}  : haloes
are more  elongated in  the central regions,  while the  outskirts are
more  rounded,  probably  due  to interactions  with  the  surrounding
environment.  Obviously the  available number  of haloes  increased in
parallel    with   computational    resources:    the   analysis    of
\citet{Jing2002} was based on  simulations with $512^{3}$ particles
in a  $100~$Mpc$\,h^{-1}$ box, which  contained hardly any halo above
$10^{14}~$M$_\odot h^{-1}$ and some higher resolution runs which provided only 12  haloes with more
than $10^{6}$ particles. On the other hand more recent works, i.e. \citet{Schneider2012}, have been able to analyse larger data sets like the Millennium    I     and     II    simulations
\citep{Springel2005a,Boylan2009}.   The  mass  range between  $10^{12}~$M$_\odot h^{-1}$  and   $10^{14}~$M$_\odot h^{-1}$  has   been  widely
explored in all these works,  while only recently small haloes down to
$10^{10}~$M$_\odot h^{-1}$  \citep{munozcuartas,Schneider2012} and some
massive haloes  of $10^{15}~$M$_\odot h^{-1}$  \citep{Despali2014} have
been  included in  this kind  of analysis.   So far,  no statistically
significant predictions are available above $3\times 10^{14}\,h^{-1}$
M$_\odot$ and we  rely on extrapolations from lower  mass haloes when it
comes to predict the shapes of massive galaxy clusters. With about 300
billion particles and a box size of $3~$Gpc$\,h^{-1}$, the Millennium
XXL simulation \citep{Angulo2012} fills the range of high
masses and explore the properties of cluster size haloes.

Our aims are twofold:
\begin{enumerate}
\item using cluster scale haloes (M$> 10^{14}~$M$_\odot h^{-1}$) from the Millennium XXL simulation, we aim to provide predictions for the shape of massive clusters.

\item then, we extend the mass range by considering haloes from the SBARBINE simulations, 
applying similar methods in order to investigate the shapes of haloes and provide
predictions over 5 decades in mass, from 
$\sim 3 \times 10^{10}~$M$_\odot h^{-1}$ to $\sim 4 \times 10^{15}~$M$_\odot h^{-1}$. 
\end{enumerate}

This paper is organized as follows: 
in Section~\ref{sec:halo_catalogue}, we present the simulations and the methodology used to 
extract haloes and measure their shapes.
In Section~\ref{sec:cluster_shapes}, we present our results for the massive cluster scale haloes,
then in Section~\ref{sec:all_shapes} we extend our analysis to a broader mass range.
In Section~\ref{sec:comparison} we compare our findings with previous works.
We discuss our results and conclude in Section~\ref{sec:conclusions}.

\section{Halo catalogue}\label{sec:halo_catalogue}
We have derived  the shape of galaxy clusters  from the Millennium XXL
Simulation (MXXL)  \citep{Angulo2012}.  To generalise  our analysis to
lower masses,  we used a new set  of simulations (Despali et  al. - in
preparation), which extended  the mass range to more than $5$ orders of
magnitudes. From both simulations we have analysed haloes from two redshifts: $z = 0$ and $z = 1$. The main features of the simulations are described in
the following sections and summarised in Table~\ref{tab_sim}.
\begin{figure*}
	\centering
	\includegraphics[width=\columnwidth]{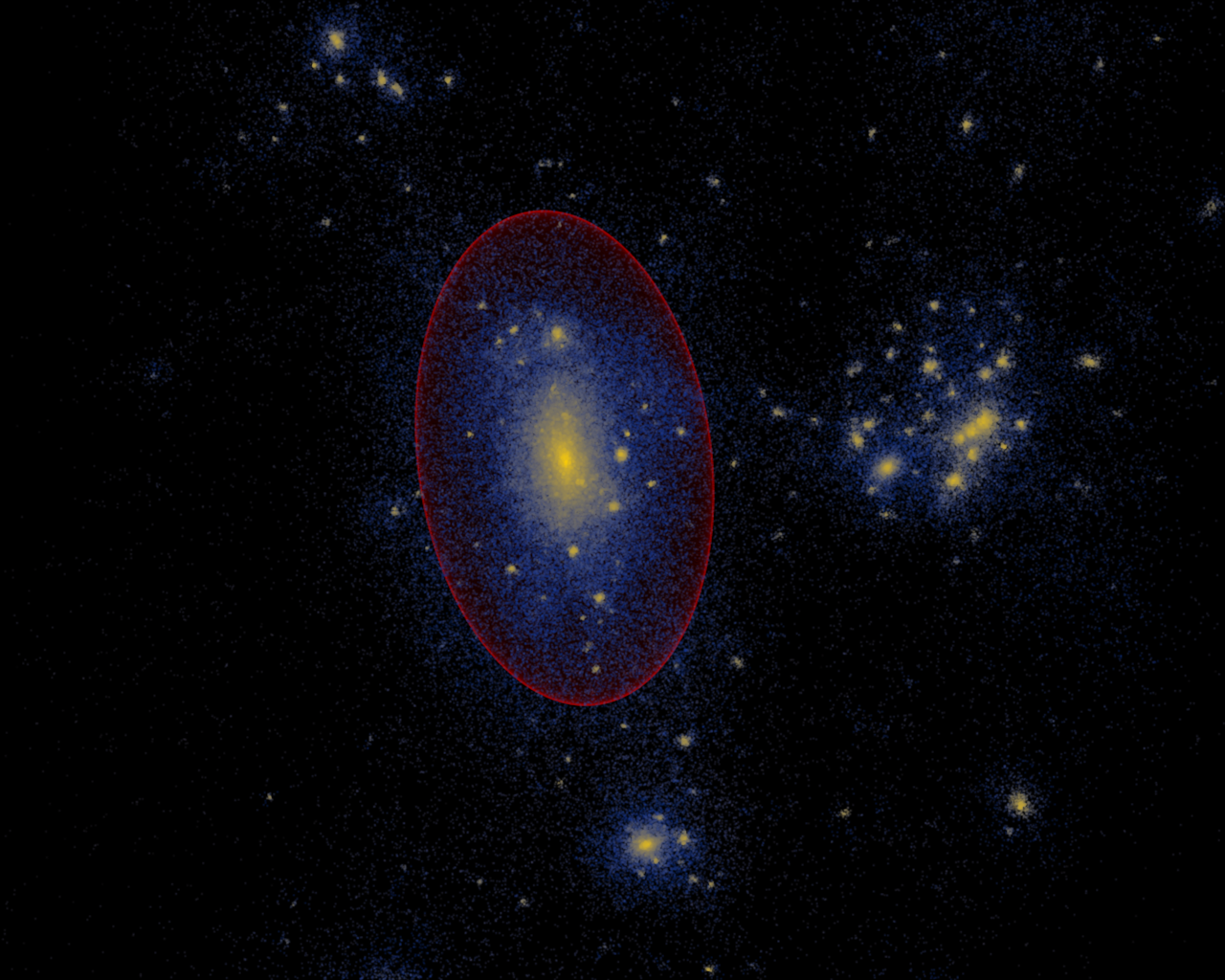} 
	\includegraphics[width=\columnwidth]{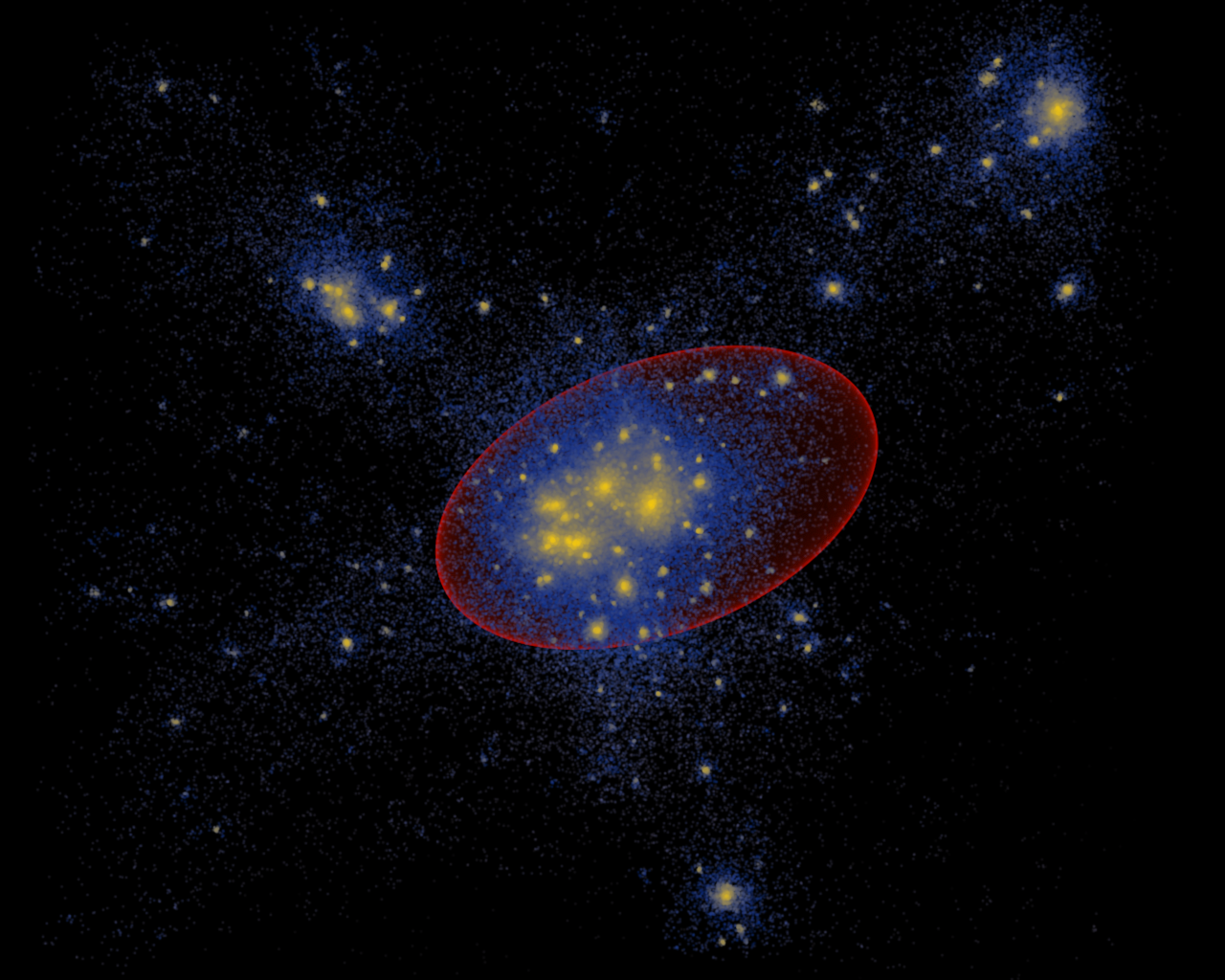} 
	\caption{Density distribution (colour scale) of dark matter particles inside a $10~$Mpc$\,h^{-1}$ side cube centred in two different
          haloes and the respective computed ellipsoids (red) that approximate the mass distribution of the halo.
          The halo shown on  the left panel  has  a   virial  mass  of  $5.29\times10^{14}~$M$_\odot h^{-1}$,   the   one  on   the   right   has   a  mass   of
          $6.90\times10^{14}~$M$_\odot h^{-1}$. These represent two families of objects: a relaxed haloes (left) and a perturbed one (right), due to the large amount of substructures the latter has to be discarded, as it can not be well described by a triaxial approximation.}
	\label{fig:haloes}
\end{figure*}
\subsection{Millennium XXL Simulation}
With  a box  side of  $3~$Gpc$\,h^{-1}$  ($4.1~$Gpc), this simulation  was
especially tailored to study massive haloes which can be only found in
very large volumes, because of their nature of extremely rare objects and due to the dampening of large fluctuation modes in smaller boxes.
The $6720^3 \approx 3 \times 10^{11}$  dark matter particles have a  mass of $6.174 \times
10^9~$M$_\odot h^{-1}$;  the   Plummer-equivalent  softening  length  is
$\epsilon =  13.7~$kpc.  For reasons  of consistency with  the previous
Millennium runs,  the adopted $\Lambda$CDM cosmology is  the WMAP one:
total matter  density $\Omega_m =  0.25$, baryons density  $\Omega_b =
0.045$,  cosmological   constant  $\Omega_\Lambda  =  0.75$,  power
spectrum  normalisation  $\sigma_8  =  0.9$ and  dimensionless  Hubble
parameter $h = 0.73$.

Due to the huge number of haloes in the simulation (almost $68$ millions at redshift $0$), we restricted the analysis to only a random sub-sample: for each logarithmic mass bin of size $0.2$ (mass inside a spherical overdensity of $200 \Omega_{crit}$) we chose either $10^5$ random objects or all, for the higher masses where the number of haloes in the bin is lower.
This cut happens at a logarithmic mass of about $14.4$ and $14.0~$M$_\odot h^{-1}$ for redshifts $0$ and $1$ respectively.
We have then re-identified haloes at redshift $z=1$ and $z=0$ using an ellipsoidal halo finder, which will be described in Section 2.3.

\subsection{LE SBARBINE simulations}

With the  purpose of comparing  different data sets and  extending the
available mass  range, we  use (from Section  \ref{sec:all_shapes} on)
the results from  five cosmological simulations which have  been run in
Padova     using    the     publicly    available     code    GADGET-2
\citep{Springel2005a}; these  are part of  a series of  new simulations
which  will  be  presented  in  a  subsequent  work  (``LE  SBARBINE''
simulations, Despali et al. - in preparation). The adopted cosmology
follows     the    recent     Planck     results    \citepalias{Planckxvi}:
$\Omega_{m}=0.307$,  $\Omega_{\Lambda}=0.693$,  $\sigma_{8}=0.829$ and
$h=0.677$.   The initial power  spectrum was  generated with  the code
CAMB \citep{Lewis2008} and initial conditions were produced perturbing
a          glass           distribution          with          N-GenIC
(\url{http://www.mpa-garching.mpg.de/gadget}).     They   all   follow
$1024^{3}$ particles in a  periodic box of variable dimension.  Haloes
were identified using a spherical overdensity algorithm \citep{Tormen2004,Giocoli2008}
 and then the best-fitting ellipsoid was found using an ellipsoidal  overdensity method,
already  presented in  \citet{Despali2013,Despali2014} and  similar to
the one used on the MXXL haloes and described in the next section; the
two codes  produce equivalent results.   We selected only  haloes with
more than  1000 particles to  ensure a good  resolution and to  have a
good comparison with the haloes of the MXXL simulation.

\begin{table*}
\centering
\begin{tabular}{|c|c|c|c|c|c|c|}
  \hline
   & box [Mpc$\,h^{-1}$] & $z_{i}$ & $m_{p}$[M$_\odot h^{-1}$] & soft
   [kpc$\,h^{-1}$] & $N_{h} (z=0)$ & $N_{rel} (z=0)$\\
  \hline
  \textbf{Ada} & 62.5 & 129 & $1.94\times 10^{7}$ &   1.5 & 39445 & 28005\\
  \textbf{Bice} & 125 & 99 & $1.55 \times 10^{8}$ & 3 & 49100 & 32107\\
  \textbf{Dora} &  500 & 99  & $9.92  \times 10^{9}$ & 12 & 66300 & 33970\\
  \textbf{Emma} &  1000 & 99  & $7.94  \times 10^{10}$ & 24 &  46665& 20696\\
  \textbf{Flora} &  2000 & 99 & $6.35  \times 10^{11}$ &  48 & 7754& 2997\\
  \textbf{MXXL} &  3000 & 63 & $6.17  \times 10^{9}$ &  18.8 & 937755 & 568477\\
  \hline
\end{tabular}
\caption{Main features of the simulations used in this work. The last
  two columns report the total number of haloes with more than 1000
  particles ($N_{h}$) and the corresponding fraction of relaxed haloes ($N_{rel}$), at redshift $z = 0$.\label{tab_sim}}
\end{table*}

\subsection{Ellispoidal halo finder}
It is known that FOF  finders tend  to  connect  together multiple  virialized
haloes  via thin  bridges  of particles  \citep{jing1994}; thus, to
characterise halo shapes more precisely, we used  a
second halo finder that iteratively selects particles inside an ellipsoid and then uses their mass distribution to compute the ellipsoid for the next step in the iteration. 

We start with a traditional spherical overdensity (SO) algorithm which
selects  particles inside a  sphere of  given overdensity,  namely the
value  from the  spherical collapse  model at  $z=0$:  $\Delta_{vir} =
359.7$ times the background  density \citep{Eke1996}, and centred in the particle with lowest potential (most bound particle).  We then compute the mass tensor\footnote{Not to be  confused with the  inertia tensor
  \citep{Bett2007}}
\begin{equation}
\mathcal{M}_{\alpha\beta} = \sum^{N_V}_{i=1} \frac{m_i \,r_{i,\alpha}\,r_{i,\beta}}{M_{TOT}}
\end{equation}
of the particles inside the virial radius  of the sphere of mass $M_{TOT}$, where
$r_i$  is the  distance of  the $i$-th  particle, of mass $m_i$, from  the  most bound
particle.
The  tensor's  eigenvectors   give  the  direction  of  the
ellipsoid  that approximate  the mass  distribution, while  the square
roots of the eigenvalues are proportional to the axes length ($c > b > a$).

Having  derived the triaxial  distribution of  dark matter  inside the
spherical overdensity,  we use it to compute  the ellipsoidal distance
of the particles  $r^2_E = x^2 + y^2 /(b/c)^2  + z^2 /(a/c)^2$, select
those inside a given radius and recompute the mass tensor with the new
subset.  We iterate this procedure  until the ratios of minor to major
axis  $s =  a/c$ and  intermediate to  major axis  $q =  b/c$ converge
within a $0.5$ per cent of error.  This algorithm has been already adopted in
the  literature \citep{Allgood2006,Schneider2012,Despali2013}, however
different  authors  use  different   values  for  the  radius  of  the
ellipsoid.  We have chosen  to follow \citet{Despali2013} who selected
the ellipsoid that  encloses an overdensity equal to  the one given by
spherical  collapse  model   $\Delta_{vir}$.   This  is  the  simplest
possible extension of the SO, which actually becomes just the first step in our
iteration, and it allows us  to adopt a more general description while
being still close to theory predictions.
As shown also by \cite{Despali2013} the difference in the measured shapes between a spherical and an ellipsoidal overdensity can not be ignored.

In figure  \ref{fig:haloes} we show  the density distribution  of dark
matter  of  two haloes (colour scale) and,  in  red,  the computed  ellipsoid  which
encloses an overdensity of $\delta_{vir}$.  The object on the left has
a virial  mass of $5.29\times10^{14}~$M$_\odot h^{-1}$ and represent a relaxed halo; the mass of the one  on the right is $6.90\times10^{14}~$M$_\odot h^{-1}$ and it is clearly multi-modal.   It can  be seen  that the ellipsoid  captures  quite well  the  overall three-dimensional  matter
distribution of the relaxed halo; though it fails, as expected, with the perturbed object.
For this reason we will need to eliminate from the catalogue all unrelaxed haloes like the one on the right panel.
A possible way to discriminate this kind of objects is to look at the offset between centre of mass and geometrical centre of the ellipsoid.
The latter is centred in the minimum of potential, which corresponds to one of the substructures; on the other hand the the centre of mass of a system like this is somewhere in between the different objects.
In turn this means that if a significant number of large substructures is present, there will be an offset between the centre of the ellipsoid and the centre of mass.

\subsection{Halo selection}
\begin{figure}
	\centering
	\includegraphics[width=\columnwidth]{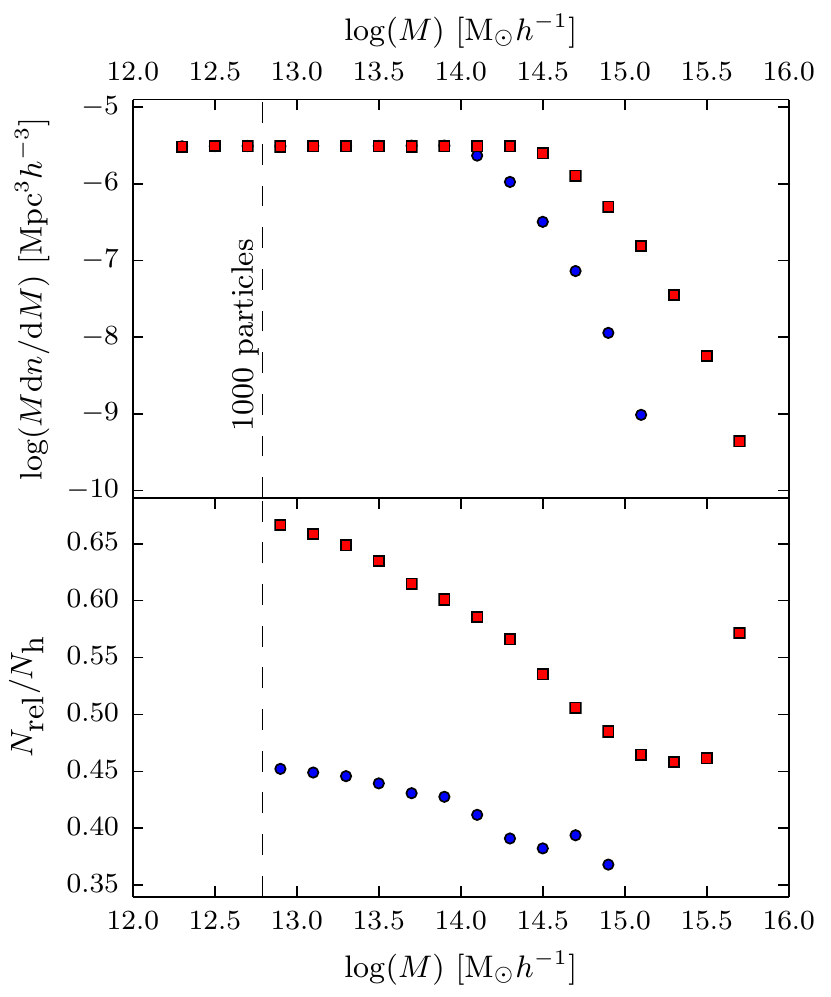}
	\caption{Mass   function and halo selection of the Millennium XXL sample at redshift $z = 0$ (red squares) and $z = 1$ (blue circles)   obtained   with   the   ellipsoidal
          overdensity. The vertical dashed line indicates the mass of an halo with $1000$ particles. \emph{Top panel}: points show the mass
          function of  the whole selected  halo catalogue. It is clearly visible the cut at low masses.\emph{Bottom
            panel}:  points  show the  percentage of  relaxed
          haloes in each mass bin  (i.e. objects with a centre offset
          smaller than $5$ per cent of their virial radius.)}
	\label{fig:mass_func}
\end{figure}
Figure \ref{fig:mass_func}  shows the mass function of  all haloes (upper panel), which is defined as the mass within the ellipsoid that encloses an overdensity of $\Delta_{vir}$, computed  by the  ellipsoidal  algorithm described above.
Data from redshift $z=0$ and $z=1$ are indicated by red squares and blue circles respectively.
As previously explained, we have analysed only a random sample of the entire halo catalogue of the MXXL simulation: this is causing the flattening at the mass bins which have more than $10^5$ objects in the entire box.
To avoid any resolution effect, we have kept only haloes with at least $1000$ particles within the ellipsoid (vertical dashed line).

Finally,  we cleaned  the halo  catalogue from  unrelaxed  haloes.  An
example of why this selection  is necessary is  the halo on  the right
panel of  Figure \ref{fig:haloes}.  The object is highly  asymmetrical and lacks of a well defined centre, therefore it can
not be described with a single triaxial model.
To remove this  effect we select only  haloes for
which the  offset between  the most bound  particle and the  centre of
mass of the particles enclosed by  the ellipsoid is less than $5$ per cent of
their virial radius:
\begin{equation}
\frac{\left| \bar{x}_{MBP} - \bar{x}_{cm} \right|}{R_{vir}} < 0.05.
\end{equation}
The lower panel of Figure \ref{fig:mass_func} shows the percentage of cleaned haloes as a function of mass.
As expected the number of perturbed haloes increases with the mass, due to more massive haloes being assembled recently.
In the past (blue circles), the percentage of relaxed haloes was lower and more constant with mass,  than at the present time (red squares).
For cluster masses the number of relaxed haloes is roughly $50$ per cent.
Although in the literature it is customary to use substructure mass fraction and virial ratio of kinetic to potential energy as measurements of the dynamical state of an halo; doing so would have required to recompute these quantities in an ellipsoidal framework.
We can compare the percentage of unrelaxed haloes with
\citet{Ludlow2012}: using a similar selection ($N_{200} > 5000$ and
spherical haloes) the fraction of objects with an offset less than $5$
per cent is $0.536$, while it becomes $0.285$ combining all the three relaxation criteria.
Therefore, by adopting this simplified selection we are still able to capture approximately $65$ per cent of all perturbed haloes.
Moreover, configurations where the offset is small but the other two criteria fails are quite symmetrical: it is possible to properly define a centre and the triaxial approximation is applicable.

This last method for selecting the relaxed haloes has also been
applied to the five SBARBINE simulations, obtaining a catalogue
equivalent to the MXXL haloes. The resulting number of relaxed haloes for both simulations at redshift $z = 0$ is shown in Table \ref{tab_sim}.

\section{Triaxial shapes of massive galaxy clusters from MXXL}\label{sec:cluster_shapes}
\subsection{Millennium XXL results}
In this first analysis we are mostly interested in the clusters mass range, therefore we will use only a portion of the available MXXL data.
By taking the ratio of minor to major axis $s = a / c$ we can measure the degree of triaxiality of a halo: the closer $s$ is to $0$, the less spherical the object is.
If we combine this information with the value of the intermediate to
major axis ratio $q = b / c$, we can infer how much prolate or oblate
the halo is.

\begin{figure}
	\centering
	\includegraphics[width=\columnwidth]{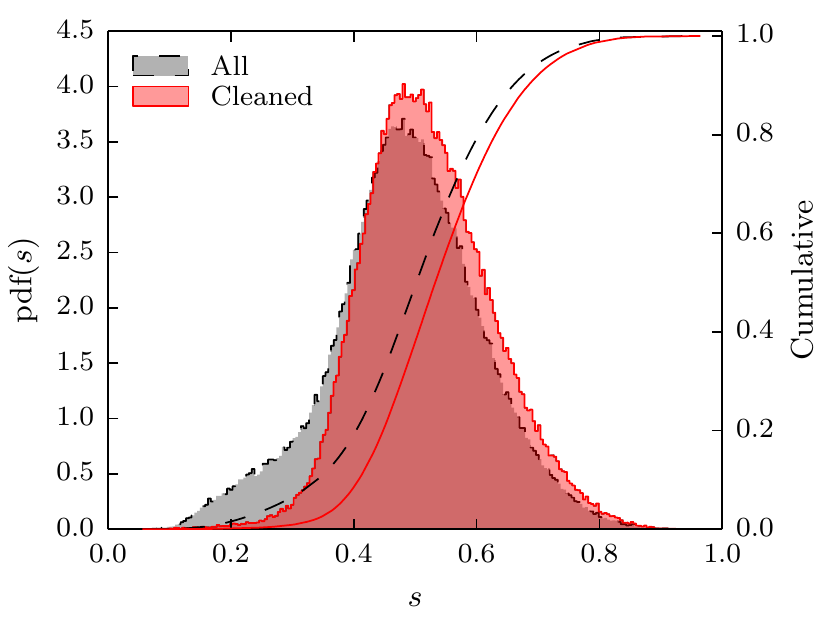}
	\caption{Probability distribution functions -- differential and
          cumulative  -- of  $s =  a /  c$. The  distributions  for the
          entire haloes  population is shown in grey  (and with dashed
          lines),  while the  red (solid)  ones refer  to  the cleaned
          population. }
	\label{fig:s_all}
\end{figure}
In Figure \ref{fig:s_all}, the distribution of $s$ is shown for the entire halo catalogue (dashed grey curves), and for the relaxed one (solid red curves).
The filled histograms represent the differential distributions, while the curves are cumulative distributions of the two different samples.
In the original population there is a noticeable bump at low $s$ which corresponds to highly aspherical objects; clearly this is the case of unrelaxed or merging clusters.
As it can be seen in the red histogram, the selection criteria we adopted have helped to remove this unwanted feature, since modelling them is beyond the goal of this work.

\begin{table}
\centering
\begin{tabular}{|c|r|c|r|c|}
   \hline
	& \multicolumn{2}{c}{$z = 0$}	&	\multicolumn{2}{c}{$z = 1$} \\
 \hline
	$\log(M) [$M$_\odot h^{-1}]$		& \multicolumn{1}{c}{$N_h$}	&	\multicolumn{1}{c}{$N_{rel}/N_h$}	& \multicolumn{1}{c}{$N_h$}	&	\multicolumn{1}{c}{$N_{rel}/N_h$} \\
  \hline
14.0 - 14.2 	& 57759 	& 58.56 \% 	& 30823 	& 41.19 \% \\
14.2 - 14.4 	& 56083 	& 56.61 \% 	& 13271 	& 39.11 \% \\
14.4 - 14.6 	& 42951 	& 53.52 \% 	& 3914 		& 38.24 \% \\
14.6 - 14.8 	& 20715 	& 50.60 \% 	& 919 		& 39.39 \% \\
14.8 - 15.0 	& 7823 		& 48.50 \% 	& 134 		& 36.81 \% \\
15.0 - 15.2 	& 2305 		& 46.46 \% 	& 6 		& 19.35 \% \\
15.2 - 15.4 	& 523 		& 45.84 \%	& &	\\
15.4 - 15.6 	& 84 		& 46.15 \%	& & \\
 \hline
\end{tabular}
\caption{Number of haloes in each logarithmic mass bin ( in $\log
  (M/M_\odot h)$ ) and percentage of relaxed haloes for redshifts $z = 0$ and $z = 1$.}
\label{tab:mass_bins}
\end{table}
We have divided our sample in eight logarithmic mass bins, from
$10^{14}~$M$_\odot h^{-1}$ to $3.98 \times 10^{15}~$M$_\odot h^{-1}$.
Table \ref{tab:mass_bins} reports the total number of
haloes $N_h$ and the percentage of relaxed ones $N_{rel}/N_h$ for each mass bin for both redshifts of the MXXL.
As expected, the number of clusters at high redshift is lower and we do not have any halo in the highest mass bins.
As noted before, the percentage of relaxed haloes is higher at low masses, which formed earlier and thus had more time to reach an equilibrium state.

It has already been established \citep{Jing2002,Allgood2006,Bett2007,Schneider2012} that the axis ratio $s$ depends on the mass of the halo, however this dependence has not been tested at the high masses available in large simulation boxes such as the Millennium XXL.
\begin{figure}
	\centering
	\includegraphics[width=\columnwidth]{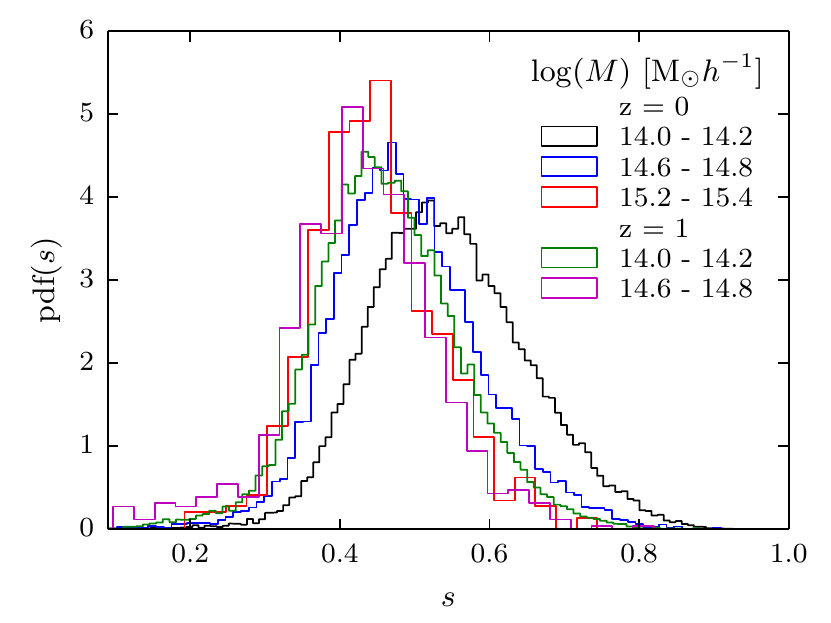}
	\caption{Probability distribution function of $s = a / c$
          binned in mass using a fixed bin of $0.2~$M$_\odot h^{-1}$ for both redshifts. For clarity, we show the results for only five of the mass bins reported in Table \ref{tab:mass_bins}.}
	\label{fig:s_masses}
\end{figure}
Figure \ref{fig:s_masses} shows the distributions of $s$ for different
masses  bins in our  sample --  only five mass bins  of Table
\ref{tab:mass_bins},  to  avoid  an  overcrowded plot;  as  halo  mass
increases, the  median value of  the axis ratio becomes  smaller, that
is, the  halo is  less spherical.
This effect is barely visible at redshift $z=1$.
Moreover  the dispersion in  $s$ is larger in the lower bins.  It is also noticeable that the distributions
are not symmetric,  particularly they are skewed to  low values of the
axis ratio.

\begin{figure}
	\centering
	\includegraphics[width=\columnwidth]{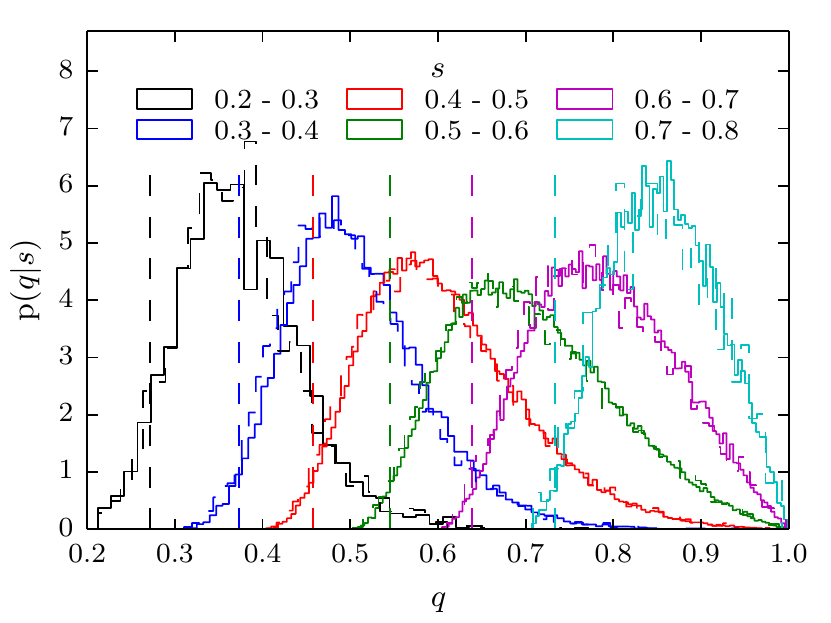}
	\caption{Conditional distributions $p(q|s)$, with $q = b /
          c$. Different colours represent the distributions for six bins
          in $s$; solid and dashed histogram shows data from redshift $0$ and $1$ respectively. The vertical dashed lines of corresponding colour show the median value of
          $s$ for each bin.}
	\label{fig:q_all}
\end{figure}
To fully describe the shape of haloes, we need also the conditional
probability distribution function $p(q|s)$, which is the distribution
of $q$ for a given value of $s$. Figure \ref{fig:q_all} shows the
conditional distributions obtained for six bins in $s$: solid histogram for $z=0$ and dashed for $z=1$.
The two redshifts are almost indistinguishable, which hints at the universality of the conditional distribution that will be discussed later on.
For any interval, the median value of $b/c$ is fairly close to the median of $a/c$ (dashed vertical lines): although still fully triaxial, haloes tend to be prolate rather than oblate.
For example, in the case of a ``disc-like'' object, all the distributions would have been prominently shifted to values close to unity, because, in this case, $b \approx c$ independently of the minor axis $a$.

\subsection{Axis ratio distribution: minor to major}\label{sec:s_mxxl}
We aim to obtain a functional form to
describe the axial ratio distributions at different masses.  Due to
the low statistic, \cite{Jing2002} were not able to fully resolve the shape of the distribution and therefore assumed a Gaussian
distribution.  On the other hand, \cite{Schneider2012} claimed to be
able to fit all the masses with a single beta distribution, although,
even after a rescaling of $s$, they mention some residual mass
dependence.  Thanks to the high statistic in the Millennium XXL simulation we are able to reconstruct
the distributions with greater detail, even at large masses.  Moreover,
we are only interested in clusters, so we do not need the
same level of generalisation of the previous authors (see section
\ref{sec:all_shapes} for broader analysis).  These two conditions
allow us to simplify the analysis and obtain a better fit of the axial
ratio distributions.

As shown by  various   authors  \citep{Press1974,Bond1991,Lacey1993}  the  mass function  written  as a  function  of peak height $\nu =  \delta_c(z)  /
\sigma(M)$  does  not  depend  on 
redshift nor  on  cosmology (see appendix \ref{sec:nu} for the details on how to compute $\nu$).
It  is easy  to  understand  why:
$\delta_c(z)$ is the critical  overdensity of the  spherical collapse
model  (initial density  required  for a  fluctuation  to collapse  at
redshift $z$), it increases  with $z$; $\sigma(M)$ is the variance in the initial density
field smoothed on  a scale of a uniform sphere of  mass $M$ and is  higher for small
masses. Then,  since in  the  past  haloes  were less  massive,  the
dependences on time of the  two quantities compensate with each other.
For example, $\nu(M_\star,z) = 1$ at  every redshift, and $\nu > 1$ always
represent a halo  with a mass larger than  the typical haloes collapsing
at  that time,  even though the  exact value  of $M_\star$  changes  with redshift.

\begin{figure}
	\centering 
	\includegraphics[width=\columnwidth]{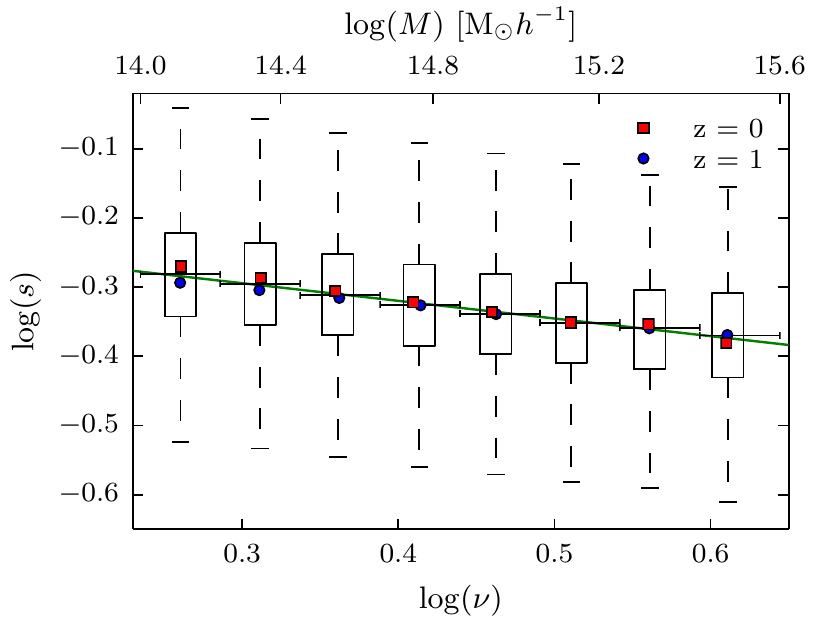}
	\caption{(logarithmic) Distribution of $s$ as  function of
          peak height: the black boxes and whiskers represent the quartiles and $1.5$ the quartiles range of the combined distributions.
		  The horizontal error shows the different bins, while the green solid line
          is the linear fit to the medians.
          Red squares and blue circles are redshift $0$ and $1$ sub-samples.}
	\label{fig:logs-logm}
\end{figure}

Figure \ref{fig:logs-logm} shows the logarithm of the first axis ratio versus logarithm of peak height ($\approx$mass) for the selected haloes.
Medians of $\log(s)$ for the two redshifts are shown in red squares and blue circles: the redshift dependence seen in fig. \ref{fig:s_masses} has disappeared completely.
As already shown by \citet[][Fig. 5]{Despali2014}, the universality of haloes properties seems to extend also to the shape when using $\nu$ instead of mass.
The change of variable allows us to provide results that are independent of the redshift and valid for different cosmologies.
This idea was already in the original \citet{Jing2002} paper, as the mass was given in units of $M_*$, but the use of $\nu$ is more general and gives a more direct connection to the theory of structure formation.
As a result, we can safely treat the two datasets as a single population, shown by the box and whiskers plot for a given $\nu$ bin (horizontal error bars).
This plot confirms the previously mentioned trend: more massive haloes (higher $\nu$) are more aspherical.

We have looked for a linear relation between $\nu$ and axis ratio in the log-log space, which translate to a relation similar to the rescaling adopted by previous authors: $
	\log\left(s\right) = a \log\left(\nu\right) + b 
	\Rightarrow \tilde{s} = 10^b = 10^{\log s -a \log\left(\nu\right)} = s \, \nu^{-a}. $
The green line is a fit of the median values; its inclination $a = -0.255 \pm 0.01$ is the opposite of the exponent in the rescaling relation and the intercept is the logarithm of the median axis ratio at $M_\star$: $\tilde{s}(M_\star) = 10^b = 0.61 \pm 0.01$, which however does not enter directly in the following relations.
The fit yields to a scaled axis ratio of:
\begin{equation}
	\tilde{s} = s \nu^{0.255};
	\label{eq:rescale}
\end{equation}
as $\nu$ takes care of any time and cosmology dependence, this rescaling is valid also for different redshifts and cosmologies.

\begin{figure}
	\centering
	\includegraphics[width=\columnwidth]{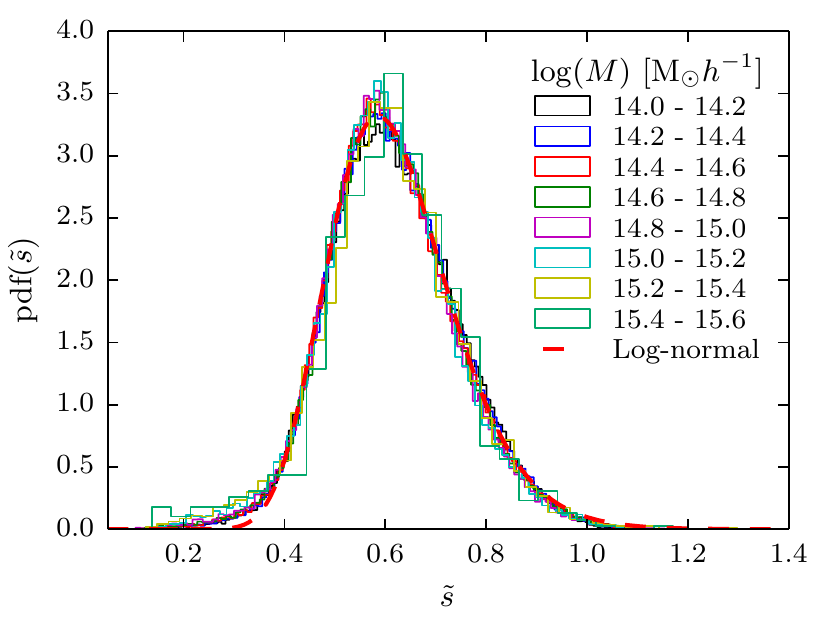}
	\caption{Distribution of the scaled axial ratio
          $\tilde{s}$ for masses shown in Table
          \ref{tab:mass_bins}. It can be easily seen that the
          distributions at all masses are well represented by an
          unique fitting function.}
	\label{fig:rescaleMXXL}
\end{figure}
As Figure \ref{fig:rescaleMXXL} shows, distributions of the rescaled axis ratios (coloured histograms) are nearly indistinguishable from each other, meaning that we have eliminated all the dependence on the mass, in contrast with the findings of \cite{Schneider2012}.
Moreover, we were not able to fit the histogram of $\tilde{s}$ with a beta distribution.
As it can be seen in Figure \ref{fig:rescaleMXXL}, the distributions are non zero at values greater than $\tilde{s} = 1$; this does not mean that there are haloes with axis ratio greater than $1$: $\tilde{s}$ is not a physical quantity, this effect is due to the rescaling.
Nevertheless, one can argue that $\tilde{s}$ represents the physical axis ratio at $\nu = 1$ ($M = 5.8 \times 10^{12}~$M$_\odot h^{-1}$); still, this rescaling has been obtained only for $M> 10^{14}~$M$_\odot h^{-1}$, leaving the unscaled axis ratio well within the physically meaningful boundaries.
We have chosen to fit the minor to major axis ratio using a log-normal distribution:
\begin{equation}
	p(x,\mu,\sigma) = \frac{1}{x\sqrt{2\pi}\sigma} exp\left( -\frac{\left( \ln x - \mu \right)^2}{2\sigma^2} \right),
\end{equation}
which corresponds to the probability distribution function of a variable which is normally distributed in the logarithmic space.
The parameters of the fitted function are the following:
\begin{equation}
\begin{split}
	\mu  = & -0.49\\
	\sigma = & \; 0.20;
\end{split}
\end{equation}
they can be converted to more familiar quantities:
\begin{equation}
\begin{split}
	\text{median }& = e^\mu = 0.61,\\
	\text{std }& = \sqrt{(e^{\sigma^2} - 1 ) e^{2\mu + \sigma^2} } = 0.13\;.
\end{split}
\end{equation}

In this framework, for a simple analysis, one can just use the scaled median value $\tilde{s} = 0.61$ with asymmetric quartiles at $0.53$ and $0.70$; then use eq. (\ref{eq:rescale}) to obtain the physical value: $s = \tilde{s} \left(\nu\right)^{-0.255}$.
On the other hand, it is possible to use the fit to obtain the whole distribution for a given mass.
For example, to use it as a prior distribution of the minor to major axis ratio, one draws a value $x$ from a normal (Gaussian) distribution with mean $\mu  = -0.49$ and standard deviation $\sigma = 0.20$, the scaled axis ratio is then $e^x$ (or directly extract $\tilde{s}$ from a lognormal distribution); inverting the rescaling relation one can obtain the axis ratio of the halo at a given peak height, which can be subsequently converted in mass for a given cosmology at a given redshift.

\subsection{Axis ratio distribution: intermediate to major}\label{sec:s-q_mxxl}
Once we are able to describe $s$ as a function of mass we can look at the correlation between the two axial ratios.
For construction, $q$ is always greater (or equal) than $s$; also it is always less than $1$.
These limits have the effect of distorting the distribution of intermediate to major axis ratio in a way that depends directly on $s$.
\begin{figure*}
	\centering
	\includegraphics[width=0.49\textwidth]{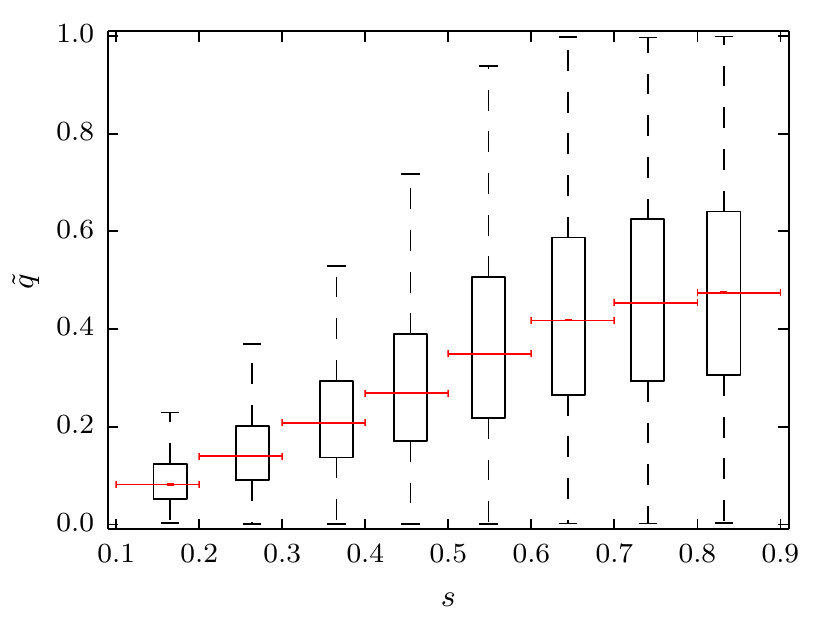}
        \includegraphics[width=0.49\textwidth]{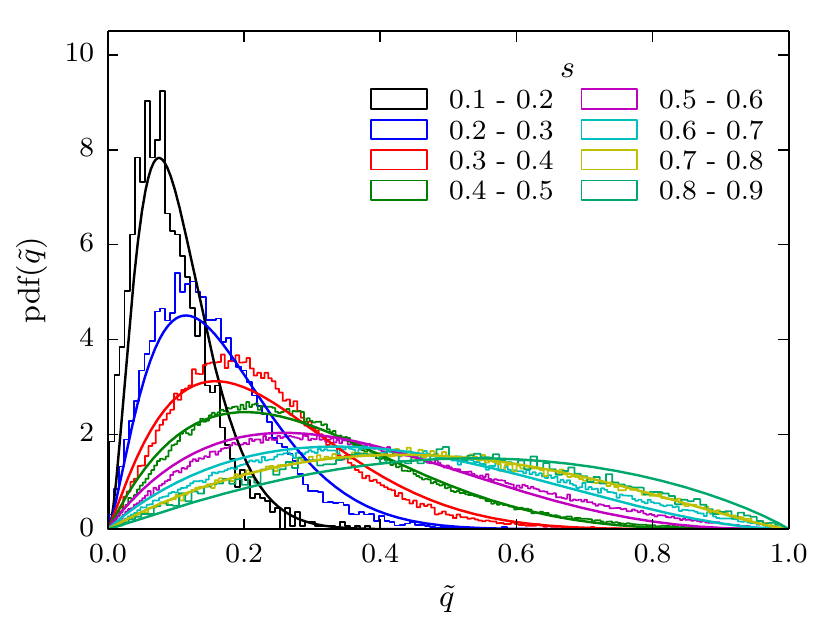}
	\caption{Distribution of $\tilde{q} = (q-s)/(1-s)$ as  function of
          $s$: the black boxes and whiskers represent the quartiles and $1.5$ the quartiles range respectively.
		  The horizontal red error bars represent the bin inside which the medians have been computed.
         $Right$ : Distributions of $\tilde{q}$ for different values of $s$ (histograms) and fitting function resulting from the model presented in the section (curves).}
	\label{fig:s-q_bins}
\end{figure*}
To avoid this problem we use the rescaled quantity $\tilde{q} = (q - s)/(1 - s)$ instead of the simple axial ratio \citep{Schneider2012}, eliminating the issues of a limited interval; the correlation between the rescaled second axial ratio and $s$ can be seen in the left panel of Figure \ref{fig:s-q_bins}, where are shown medians (red error bars) and quartiles (box and whiskers plot) for different values of the first axis ratio.
We have divided $\tilde{q}$ in bins of different $s$ and extracted the distributions $p(\tilde{q}|s)$ (right panel of Fig. \ref{fig:s-q_bins}).
From both plots, it is quite evident that $\tilde{q}$ strongly depends on the first axial ratio, with higher values at higher $s$, which is in agreement with haloes that tend to be prolate.
Moreover the scatter is larger at higher $s$, though this is mostly due to the rescaling which extends the allowed interval of $\tilde{q}$.

Because of the strong correlation between $\tilde{q}$ and $s$, we can not just give $\tilde{q}$ as a function of mass: to obtain the second axis ratio distribution for a given mass, we have to describe $p(\tilde{q}|s)$ and then get the first axis ratio from its distribution at different masses (as shown in Section \ref{sec:s_mxxl}).
Given the large differences in the shapes of the distributions of $\tilde{q}$ at a given $s$, the rescaling needed to reduce them to a single one needs to be much more complex than the one adopted in the last section.
Therefore, we fit each single histogram with a different beta distribution, which has the following analytical expression:
\begin{equation}
	p(x,\alpha,\beta) = \frac{1}{B(\alpha,\beta)} x^{\alpha-1}(1-x)^{\beta-1}.
\end{equation}
This function has two shape parameters $\alpha$ and $\beta$; the factor $1/B(\alpha,\beta)$ is a normalisation constant that can be computed by requiring that the integral of the probability distribution function is equal to unity.

From the fitting procedure we obtained a pair of parameters for each bin in $s$; however $\alpha$ has a complicated dependence on the first axial ratio (almost constant with an average value of $\alpha = 2.15$), while the mean value of the beta distributions $\mu = 1 / (1 + \beta/\alpha)$ follows a linear relation.
\begin{figure}
	\centering
	\includegraphics[width=\columnwidth]{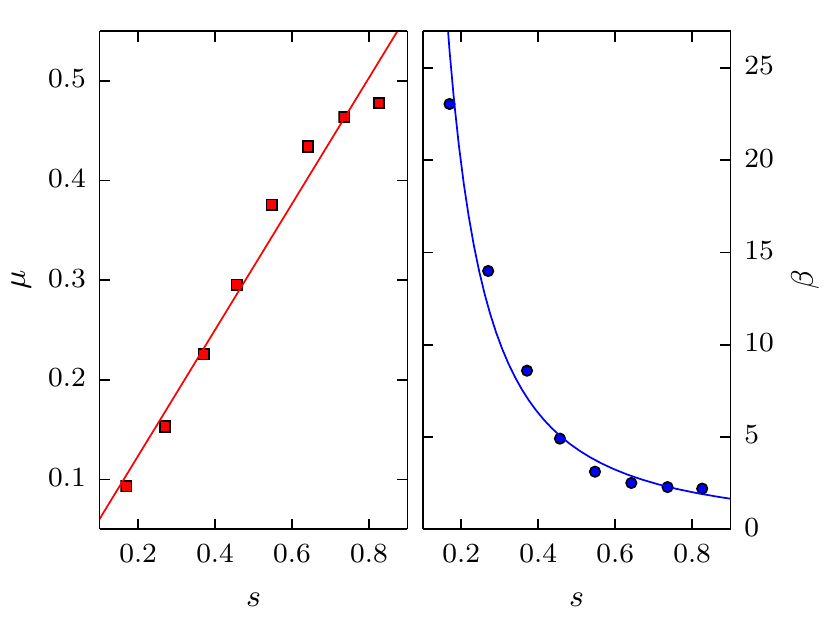}
	\caption{Parameters of the fitted beta functions. Red is the mean of the distribution, in blue the second parameter $\beta$.}
	\label{fig:q1_parameters}
\end{figure}
Figure \ref{fig:q1_parameters} shows the dependence of the mean $\mu$ (red squares on left panel) and $\beta$ parameter (blue circles on right panel) of the fitted beta functions on the first axial ratio $s$.
The coloured lines in each respective panel show a fit of these two parameters:
\begin{equation}
\begin{split}
	\mu(s) 		& = 0.633 s -0.007 \\
	\beta(s)	& = 1.389 s^{-1.685} .
\end{split}
\end{equation}
These two equations give us a functional form of $p(\tilde{q}|s)$: starting from a value of $s$, one can retrieve the mean $\mu$ and $\beta$ from which the other parameter can be computed $\alpha = \beta/(1/\mu + 1)$.
This gives what is needed to reconstruct the distribution of $\tilde{q}$ of a given $s$ and the scatter, if needed.
The final step is to revert the change of coordinates and compute the physical axial ratio $q$.

\section{Exploding the mass range to 5 orders of magnitude}\label{sec:all_shapes}
The next  step of our  work is to  explode the recipes for  dark matter
halo shapes to  lower masses: in the following  sections we describe
how to generalise the axial  ratio distribution to a wider mass range.
To do so,  we combined the MXXL data with  the SBARBINE simulations, a
set of cosmological simulations that  will allow us to study the shape
of  dark  matter  haloes  from  $3\times10^{10}~$M$_\odot h^{-1}$  to
$6\times10^{15}~$M$_\odot h^{-1}$.

As before, we express the mass dependence in terms of peak height $\nu$.
By doing this, it is possible to treat homogeneously data from different redshifts and cosmologies, such as the SBARBINE and the MXXL simulations.

\subsection{Axis ratio distribution: minor to major}\label{sec:s_all}
\begin{figure*}
	\centering
	\includegraphics[width=0.49\textwidth]{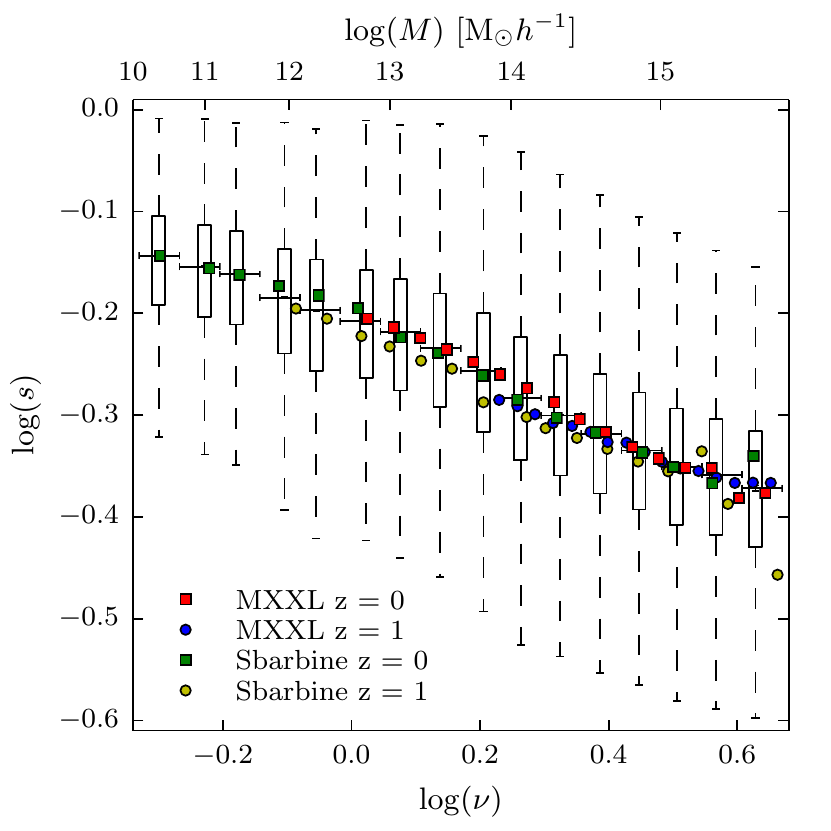}
        \includegraphics[width=0.49\textwidth]{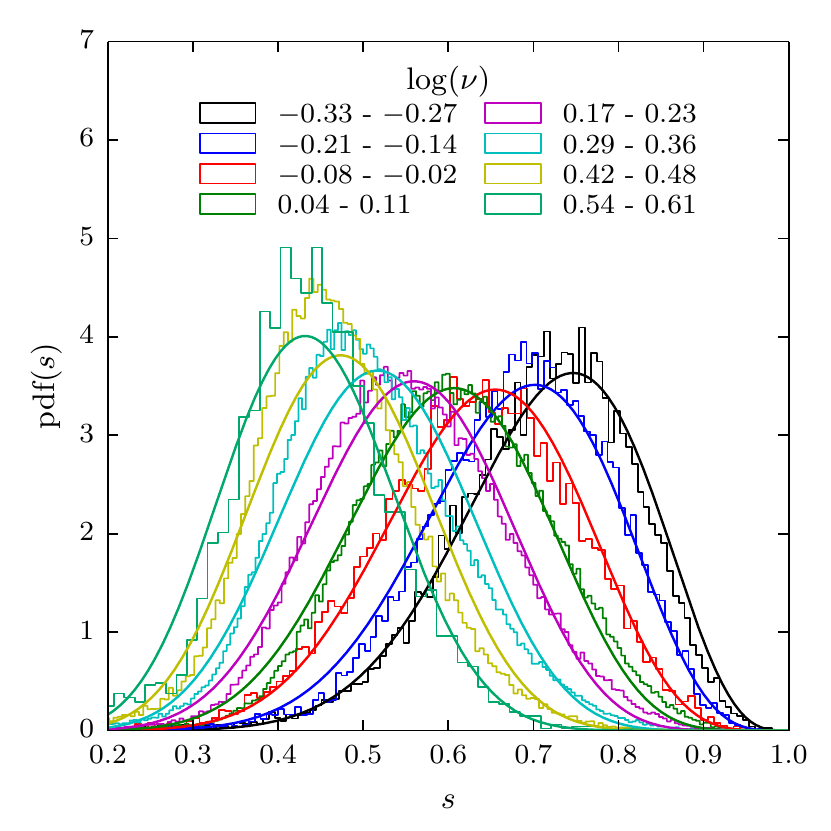}
	\caption{ $Left$: Distribution of $s$ as function of
          peak height for all the haloes selected from both redshifts the two
          simulations: the black boxes and whiskers represent the quartiles and $1.5$ the quartiles range respectively computed within the bins shown by the horizontal error bars. The coloured points represent the medians for individual redshifts for the two simulations.
          $Right$ : Differential distribution of $s$ for 8 bin in
          $\nu$ (histograms) and the respective approximating functions obtained as shown in the section (curves).}
	\label{fig:s-nu_bins}
\end{figure*}
On left panel of figure  \ref{fig:s-nu_bins}, the logarithm of the minor to major axial
ratio  $s$ is shown  as a function of  the  logarithm  of $\nu$. As before, horizontal error bars represent the interval in $\nu$ and the box and whiskers are the quartiles and 1.5 the quartiles range for the combined sample, while coloured points are medians of individual catalogues.
Again, there is no difference in the medians between redshifts, neither between the single simulations.
It can  be seen that $s$  has a nearly
linear dependence  on $\log(\nu)$, with  a hint of flattening  at both
high and low masses. 

For each bin, we extracted  the probability distribution function of $\log(s)$
(right  panel  of  Fig.  \ref{fig:s-nu_bins}).  The  resulting  curves
exhibit  an interesting  pattern: high  and low  $\nu$  histograms are
mirrored  with respect to  a central  symmetric distribution  which corresponds to $\nu = 1.21$ ($M \approx M_*$).
The rescaling adopted in section \ref{sec:s_mxxl}  does not compensate  this large variation in the form of the distributions and it is not  able to remove entirely
the mass dependence.  Instead  of using a different rescaling relation
to obtain a  single pdf, we decided to follow the  same recipe we used
for  the second  axial  ratio: first  of  all we  separately fit  each
distribution  and  then we  relate  the  resulting  parameters to  the
binning quantity.
\begin{figure}
	\centering
	\includegraphics[width=\columnwidth]{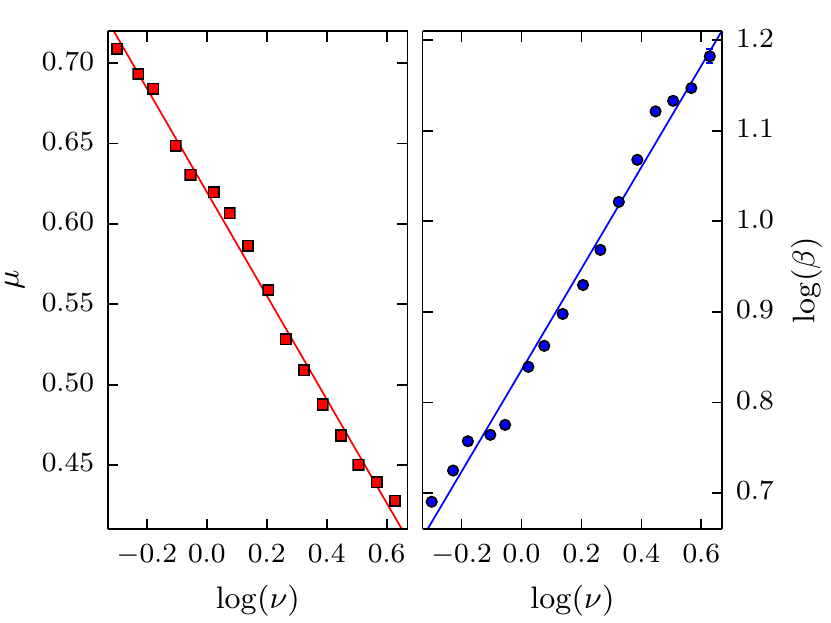}
	\caption{Parameters of the fitted beta functions. Red is the mean of the distribution, in blue the second parameter $\beta$.}
	\label{fig:s_parameters}
\end{figure}
This is shown in Figure  \ref{fig:s_parameters}, where we fit the mean
(left  panel)  and  $\beta$   parameter  (right  panel)  of  the  Beta
distributions    we   derived   by    fitting   the    histograms   of
the right panel of Fig. \ref{fig:s-nu_bins}. In order to keep the procedure simple we fit with a linear relation both $\mu$ and $\log \beta$:
\begin{equation}
\begin{split}
	\mu(\nu) 		& = - 0.322\log \nu + 0.620 \\
	\log\left(\beta(\nu) \right)		& = 0.560\log \nu + 0.836.
\end{split}
\end{equation}
As before, the dependence of $\alpha$ is difficult to describe and it is almost constant with a value of about $11.21$.

Using  this  fits we  are  now  able  to approximate  the  probability
distribution function  of the first  axial ratio with a  Beta function
with parameters $\alpha = \beta/(1/\mu  -1)$ and $\beta$, over a range
in mass of almost $6$ orders of magnitudes.   Moreover the use of
$\nu$ allows  us to extend  these results to different  cosmologies and
different redshifts.

\subsection{Axis ratio distribution: intermediate to major}
\begin{figure}
	\centering
	\includegraphics[width=\columnwidth]{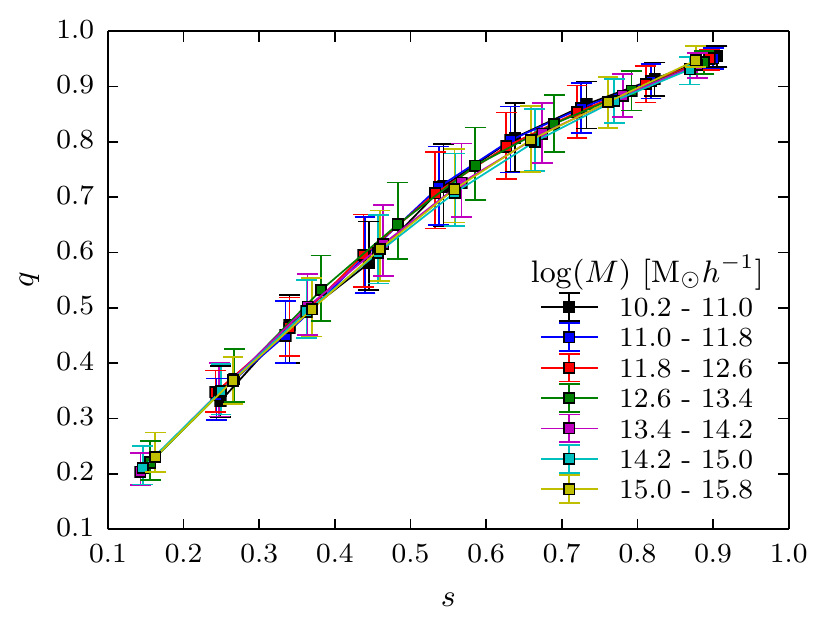}
	\caption{Axis ratio $q$ as function of $s$ for different
          masses, represented by the points of different
          colours. Since there is no residual mass dependence in the
          conditional distribution, we get the same result as in the
          MXXL with all the simulations, confirming that this relation
        is universal.}
	\label{fig:s-q_masses}
\end{figure}
Finally,  to  fully describe  a  triaxial halo  of  a  given mass  the
intermediate to  major axis ratio  has to be parametrized.   As Figure
\ref{fig:s-q_masses} shows, the relation  between $q$ and $s$ at redshift $z=0$ does not
depend on the  mass: the curves of different  colours represent
different mass bins and still trace the same relation.  The fact that all the mass
dependence is already inside $s$, allows us to use for $p(q|s)$  the same functional
form of  section \ref{sec:s-q_mxxl}, independently of the
mass we choose.
The same applies to different redshifts (not shown here, but see \ref{fig:q_all} for a limited comparison), with the relation  between the two quantities being indistinguishable from the one in Fig. \ref{fig:s-q_masses}.
Moreover, this independence of the conditional distribution from both mass and redshift is in agreement with the theoretical predictions from \citet{Rossi2011}.

\section{Comparison with previous works}\label{sec:comparison}
\begin{figure*}
	\centering
	\includegraphics[width=0.7\textwidth]{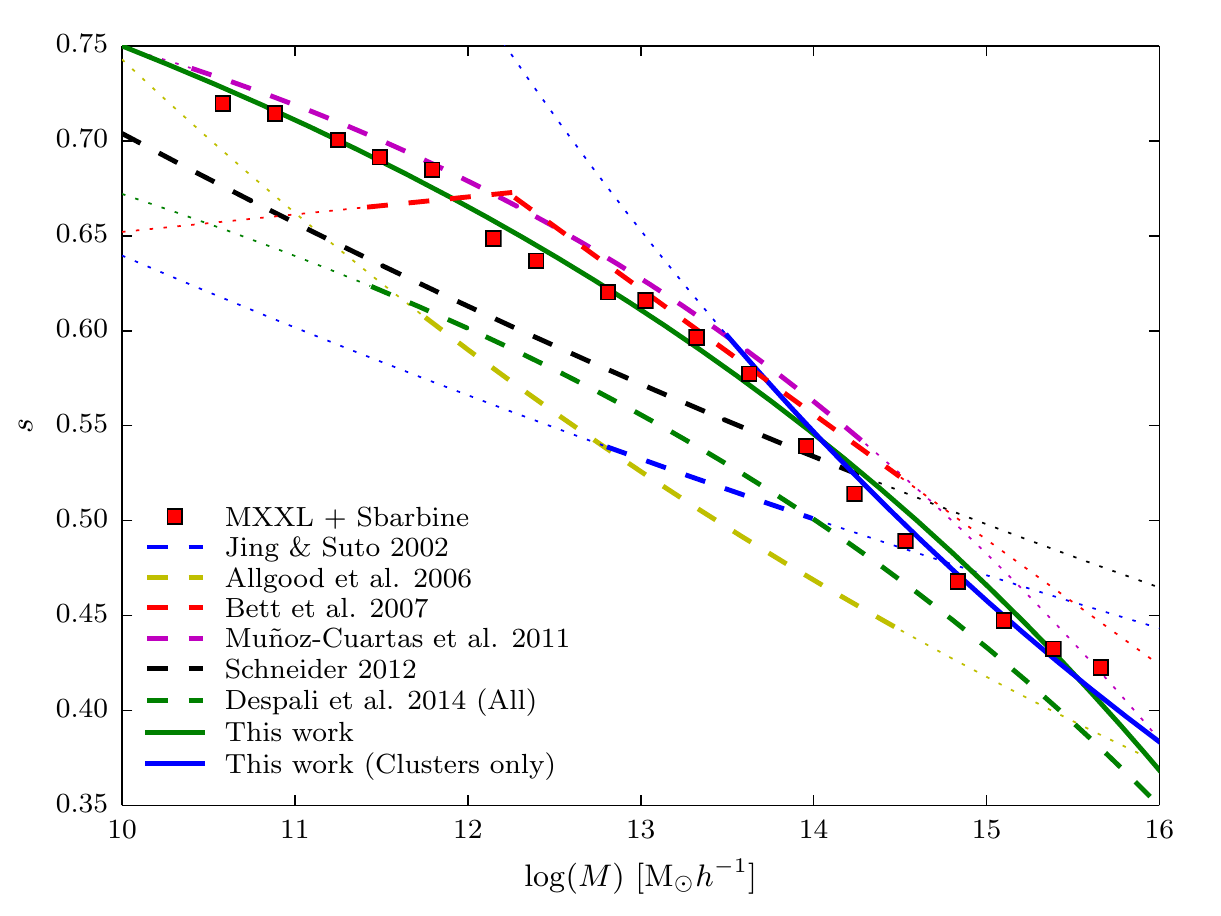}
	\caption{Comparison between previous works (dashed lines) and the results of this paper (solid lines). Red squares represent the data from both redshifts of the MXXL and the SBARBINE simulations, converted to redshift $z=0$ for the Millennium cosmology. The blue solid line is the model for clusters shown in section \ref{sec:s_mxxl}; the green solid line is the fit for the entire mass interval from section \ref{sec:s_all}. The dotted parts of the curves show the mass ranges outside where the relations have been derived from.}
	\label{fig:comparison_s}
\end{figure*}
We  have compared  our results  with measurement  of axis  ratios from
other  authors  (Fig.  \ref{fig:comparison_s}).   The data  from  both
redshifts of the Millennium XXL  and SBARBINE simulations  are shown with  red squares,
the   median   result   form    the   analysis   on   cluster   masses
(sec. \ref{sec:s_mxxl})  is the  blue solid line  and the  green solid
line is  from the  combined datasets (sec.  \ref{sec:s_all}).  Results
from other authors are shown with dashed lines in the mass range where
their analysis was carried out and with dotted lines when extrapolated
beyond it. Moreover all data and predictions have been converted to redshift $z=0$ for the Millennium cosmology, when necessary.  As it can  be seen, there  is a general agreement  in the
dependence of  $s$ on  the mass, with  more massive haloes  being less
spherical.  Although there seems to be a scatter of about $15$ per cent, this is
due  more  to  the  differences  in the  method  of  measuring  shapes
(different finders, radius, cleaning  procedure), than an error on the
measurement.
It must be noticed that instead of the spherical mass, we used the mass within the ellipsoid for consistency reasons; yet, this does not substantially alter the findings presented here.

The most important difference comes from the radius at which the shape
is measured.  \citet{Jing2002} (blue  dashed line) used particles of
the isodensity  surface corresponding to $2500\delta_c$,  roughly at a
radius of $0.3 R_\text{vir}$; this  analysis is different from all the
following authors, as  it reflects the shape of  an ellipsoidal shell,
and  not of  all  the mass  inside  the ellipsoid.   Their mass  range
$6\times10^{12}$ -- $10^{14}~$M$_\odot h^{-1}$ was also quite small compared
to later analysis.

Studying  a  larger mass  interval,  $6\times10^{11}$ --  $3\times10^{14}~$M$_\odot h^{-1}$, \citet{Allgood2006}  (yellow dashed line)  derived axis
ratios    of   particles    distribution    inside   $0.3R_\text{vir}$
diagonalizing  the normalised  mass tensor  (weighted by  the distance
from the centre); because of  this their measure reflects the shape at
a even closer radius.

On the other hand,  \citet{Schneider2012} (black dashed line) extended
the  analysis up to  the virial  radius, nevertheless  the use  of the
normalised tensor prevents a meaningful comparison with our results.

All of  these results  are lower  than what we  derived, which  can be
explained by  the fact that the  shapes were measured  at inner radii,
where the particle distribution is supposed to be more elongate .
However, if  we restrict the  comparison to works that  used particles
within the virial radius the  agreement becomes much more strong. This
is the case of \citet{munozcuartas} (magenta  dashed  line), who studied  shapes
with an  ellipsoidal overdensity algorithm similar to  the one adopted
in this work: their results agree with ours much more than any other work.

Finally, using a different  type of halo finder, \citet{Bett2007} (red
dashed line)  measured $s$ for a  set of particles  that represent all
the  bounded particles  of  an halo  without  assuming any  particular
shape; the  finder also  clean the sample  for unrelaxed  haloes.  The
agreement with our results is  another indication that an ellipsoid is
a good approximation for relaxed haloes.

The other  difference can arise from  the cleaning of  the sample; the
green dashed  line show the prediction  from \citet{Despali2014}, which
is obtained from all haloes,  regardless of their state of relaxation.
As expected the values are  lower, since unrelaxed haloes are typically
irregular  and so they appear  more  elongated with  lower  axial ratios.  The
difference is greater for less  massive haloes.

\section{Summary and Conclusions}\label{sec:conclusions}
We  have  studied the  triaxiality  of  dark  matter haloes  from  the
Millennium XXL Simulation, which  enabled us to characterise the shape
of haloes  with extremely good  statistic in the galaxy  clusters mass
range,  from  $10^{14}~$M$_\odot h^{-1}$  to  $4\times  10^{15}~$M$_\odot h^{-1}$.  Using the SBARBINE  simulations, we have 
extended our analysis to  lower masses down to $3\times10^{10}~$M$_\odot h^{-1}$, thus increased the mass range by more than 5 orders of magnitude.
The main results of our analysis are the following:
\begin{itemize}
\item dark matter haloes are triaxial with a tendency of being prolate
   and  in particular  more massive  objects  are less
  spherical; as shown in Fig. \ref{fig:s_all} unrelaxed haloes have the effect of artificially
  increasing  the axis  ratios and  can not  be  described  by this
  simple ellipsoidal model, which is unimodal by construction;
\item for  clusters, the distribution  of the rescaled minor  to major
  axis  ratio  is  well  described  by a  lognormal  distribution,  in
  contrast to  previous extrapolations from lower masses  that found a
  simple Gaussian fit;
\item over the whole examined mass range, $s$ can be approximated by
          a  beta distribution that  depends only  on the  peak height
          $\nu$;
\item the conditional  intermediate  to major  axis
          ratio distribution $p(q|s)$ can  also be described by a beta
          distribution that  depends only on the first  axis ratio and not on the mass, thus the same approach can be used for both
          clusters and the whole mass range of haloes;
\item overall, the probability distribution  function of the shape of a dark matter halo is given by one single parameter $\nu$, related to its mass, that incorporates the dependence   on  redshift   and cosmology.
This goes in support of methods that allows to change the cosmology of a numerical simulation \citep{Angulo2010}, as within good approximation most of halo properties depend only on $\nu$.
\end{itemize}

In the recipe that we provide, an halo shape is determined only by its mass and can be changed to different cosmologies and redshifts.
Depending on the level of precision desired, it is possible to choose different approximations:
\begin{itemize}
\item for a simpler analysis that is focused on the entire mass range, section \ref{sec:s_all} presents a single method that can be applied to masses from $10^{10}$ up to $10^{16}~$M$_\odot h^{-1}$.
If restricted to masses lower than $10^{14}~$M$_\odot h^{-1}$, this is actually a very accurate description of haloes shapes;
\item if the interest is only on clusters shapes, then section \ref{sec:s_mxxl} gives a more precise model;
\item finally, it is possible to combine the two description and just use the most suitable one given the mass of the halo, although losing the universality of the description.
\end{itemize}
A simple implementation of this model can be found on a dedicated website\footnote{\href{http://people.lam.fr/bonamigo.mario/triaxial/}{http://people.lam.fr/bonamigo.mario/triaxial/}}.

In  section \ref{sec:comparison}  we  have compared  our results  with
previous findings.   There is a general agreement  with previous works
within a $15$ per cent scatter that is  due to the different methods used and
especially to the radius at which the shape is measured.  However, the
picture is clear: dark matter  haloes are triaxial objects and this effect is more prominent in clusters where  the spherical  model is
quite  far  from being  able  to  realistically  represent the  matter
distribution.

\section*{Acknowledgments}
This work  has been  carried out  thanks to the  support of  the OCEVU
Labex (ANR-11-LABX-0060) and the A*MIDEX project (ANR-11-IDEX-0001-02)
funded  by the  "Investissements d'Avenir"  French  government program
managed  by  the ANR.   ML  acknowledges  the  Centre National  de  la
Recherche Scientifique  (CNRS) for its support. 
This study also benefited from the facilities offered by CeSAM (Centre de donn\'eeS Astrophysiques de Marseille \href{http://lam.oamp.fr/cesam/}{http://lam.oamp.fr/cesam/}).
GD  has been partially
financed by the the Strategic Research Project $AACSE$ (Algorithms and
Architectures  for  Computational  Science  and  Engineering)  of  the
University of  Padova.  CG's research  is part of the  project GLENCO,
funded under  the European  Seventh Framework Programme,  Ideas, Grant
Agreement n.  259349. CG and  RA thank LAM for supporting their visits
during  which part  of this  work has  been done.  GD and CG thank  the whole
cosmology group  of the  University of Padova  with whom the SBARBINE
simulations were designed: in  particular we thank Giuseppe Tormen for
providing  the computational  resources and  Giacomo Baso  for running
$Ada$.

\bibliographystyle{mn2e}
\bibliography{./library.bib}

\appendix 
\section{Density peak height}\label{sec:nu}
In this appendix we describe step  by step how to compute density peak
height $\nu$ for a virialized halo with mass $M$ at redshift $z$ for a
given cosmological model.  Its definition is the following:
\begin{equation}
\nu \equiv \frac{\delta_c(z)}{\sigma(M)},
\end{equation}
where  $\delta_c(z)$  is the  critical  overdensity  of the  spherical
collapse  model,  the  initial  density   required  for  a
fluctuation  to  collapse  at  redshift  $z$.  This  in  turn  can  be
expressed as the collapse overdensity  at redshift $z=0$ rescaled to a
given time:  $\delta_c(z) =  \delta_c / D(z)$,  with $D(z)$  being the
linear growth rate of a density fluctuation normalised to unity at the
present time. The overdensity $\delta_c$ depends only on redshift and not on the mass; on the other  hand, the denominator  $\sigma(M)$, depends on the mass but not on redshift.  It  is the  variance in the  initial density
field smoothed on a linear scale  $R$, which corresponds to the radius
of a  uniform sphere of mass  $M$.  Therefore, only the  linear growth
rate $D(z)$ and the initial power spectrum $P(k)$ are needed.

From the linear perturbation theory, it is possible to compute $D(z)$:
\begin{equation}
	D(z) \propto H(t) \int_0^t \frac{\text{d}t'}{a^2(t')H^2(t')},
\end{equation}
which has to be solved numerically.
Fortunately, there is an approximated solution \citep{Carroll1992} that can be expressed as $D(z) \propto g(z) / (1+z)$, where:
\begin{equation}
	g(z) = \frac{5/2\, \Omega_m(z)}{\Omega_m^{4/7} - \Omega_\Lambda(z) + \left[1+\Omega_m(z)/2 \right] \left[ 1+\Omega_\Lambda(z)/70 \right] }.
	\label{eq:growingmode}
\end{equation}
Additionally,  the   collapse  overdensity   has  an   extremely  weak
dependence on  cosmology: $\delta_c \approx  1.686\left[\Omega_m (t_c)
  \right]^{0.0055}$;   for   realistic   cosmologies   this   can   be
approximated to  $\delta_c \approx  1.69$.  Therefore, at  the present
time  the collapse  overdensity is  $\delta_c$ and  it increases  with
redshift, due to $D(z)$.

The other quantity required, the variance $\sigma^2(M)$, is defined from the power spectrum as:
\begin{equation}
	\sigma^2(M) = \frac{1}{2\pi^2} \int_0^\infty P(k) \tilde{W}^2(kR) k^2 \text{d} k;
	\label{eq:sigma}
\end{equation}
where $\tilde{W}$ is the Fourier transform of a window function.
Typically, $W$ is a Top Hat (sphere) in the coordinates space, so that its Fourier transform $\tilde{W}$ is:
\begin{equation}
	\tilde{W}(kR) = 3\frac{\sin\left(kR\right) - kR \cos\left(kR\right)}{\left(kR\right)^3};
\end{equation}
with the  radius $R$  given by $M  = \rho_b 4  \pi/3 R^3$.   The power
spectrum $P(k)$ of the density fluctuations is the main input; given a
set of cosmological parameters it can  be computed from a software like
CAMB \citep{Lewis2008}.  As it is function of initial conditions only,
$\sigma(M)$ needs to be computed only  once for a given cosmology: all
the redshift dependence is inside $D(z)$.

Finally, for  an halo of  mass $M$,  using eq. (\ref{eq:sigma})  it is
possible  to compute  $\sigma(M)$ and  combine  it with  the value  of
$D(z)$ from  eq. (\ref{eq:growingmode}) to obtain  the correct density
peak height~$\nu$.
\label{lastpage}
\end{document}